# On the segregation of Re at dislocations in the γ' phase of Ni-based single crystal superalloys

Xiaoxiang Wu[1,2]*, Surendra Kumar Makineni[1]*, Paraskevas Kontis[1], Gerhard Dehm[1], Dierk Raabe[1], Baptiste Gault[1]*, Gunther Eggeler[2]*

[1] Max-Planck-Institut für Eisenforschung GmbH, 40237 Düsseldorf, Germany

[2] Institut für Werkstoffe, Ruhr-Universität Bochum, Universitätsstrasse 150, D-44 780 Bochum, Germany

*Correspondence to: x.wu@mpie.de (XW), sk.makineni@mpie.de (SKM), b.gault@mpie.de (BG), gunther.eggeler@ruhr-uni-bochum.de (GE)

**Abstract**

We report evidence of Re and Mo segregation (up to 2.6 at.% and 1 at.%) along with Cr and Co to the dislocations inside of γ' precipitates in a second generation Ni-based single crystal superalloy, after creep deformation at 750°C under an applied stress of 800 MPa. The observed segregation effects can be rationalized through bridging the solute partitioning behavior across the γ/γ' interface and the pipe diffusion mechanism along the core of the dislocation line. This understanding can provide new insights enabling improved alloy design.

Single crystal Ni-based superalloys have long been important engineering materials used in gas turbines and jet engines due to their superior creep, fatigue and oxidation properties at elevated temperatures [1–7]. These are associated with the two-phase microstructure, where a disordered face-centered cubic (FCC) γ matrix contains a high volume fraction of $L1_2$ ordered γ' precipitates, which have been considered initially as dislocation-free [3]. Many elements are typically added for solid solution strengthening of the γ matrix, such as W and Mo [8]. Re has attracted special attention due to the fact that additions of 2 to 6 wt.% Re significantly enhance the creep properties of Ni-based single crystal superalloys [1,3,9–11]. How Re improves the creep properties is debated. The following possibilities have been suggested in the literature: (i) solid solution strengthening and distortions of the γ matrix lattice due to the large size of Re atoms [1,12], (ii) the slow diffusion rate of Re in Ni (slowest of all d-shell elements) [13,14], and (iii) the possible formation of Re clusters in the γ matrix that might hinder dislocation movement [15–18]. However, these Re clusters were neither detected by extended X-ray absorption fine structure investigations (EXAFS) nor by atom probe tomography (APT) [12,19,20]. Recent experiments show that Re atoms segregate to the core of network dislocations at γ/γ' interfaces and hence block their motion during creep deformation [21,22]. Till now, the observation of effects associated to Re has been confined either in the γ matrix phase or at the γ/γ' interfaces.

Elements such as W, Re, Co and Cr typically partition to the γ matrix in single crystal Ni-based superalloys. In contrast, Al, Ti and Ta favor the formation of γ' particles and are observed to partition to γ' [3,6,23–26]. Similarly, at a smaller scale, segregation of certain solute elements to dislocations (Cottrell atmosphere) and stacking faults (Suzuki effect) has been observed in Ni-based superalloys [27–29]. Viswanathan et al. [30] were the first to report segregation of Cr and Co to a superlattice-intrinsic stacking fault (SISF) inside γ' in two commercial Ni-based superalloys, using energy dispersive spectroscopy (EDS) in the scanning transmission electron microscopy (STEM) mode. Smith et al. [31,32] extended the investigations and tackled the segregation behavior to superlattice-extrinsic stacking fault (SESF) and found Nb, Ta, Ti and Co enrichment with local structural phase transformation from $L1_2$ (γ' phase) to a $DO_{24}$ ordering (η phase). Similar segregation behavior has been found at dislocations [33–37], twin boundaries [38–40] and planar defects of γ' phase in Co-based superalloys (enrichment of Co and W in SISF and of Co and Cr in anti-phase boundary) [37,41,42]. This raised the question whether these elementary segregation processes are associated to the rate limiting steps for creep deformation.

Here, we show evidence by APT of Re and Mo segregation, along with Cr and Co, to dislocations present inside of γ' particles. We show that these segregation effects to the dislocations are broadly related to chemical partitioning, which can contribute to control the creep rate.

Here we investigated a creep deformed, second generation single crystal Ni-based superalloy referred to as ERBO1, which is akin to the CMSX-4 alloy, containing ~ 3wt.% Re. More information about the material can be found in ref. [6]. A series of interrupted low temperature (750°C) high stress (800 MPa) tensile creep experiments were conducted along precisely determined [001] orientations. The details of how strain rates evolve with time are reported in refs. [43,44]. For the present work, two creep stages (i.e. stage II:1% and stage III:5%) were selected for precise structural and chemical analysis using TEM and APT. The 1% specimen was just about to pass through a local intermediate maximum of creep rate (~3h, $1.6 \times 10^{-6}$ s$^{-1}$), while the 5% specimen had reached a global creep rate minimum (~172h, $3.5\times10^{-8}$ s$^{-1}$). TEM foils were prepared from the gauge area of the crept samples, and cut out parallel to the {111} planes for characterization of elongated dislocation segments and extended planar defects. More details about the TEM foil preparation can be found in ref. [45]. The microstructure was investigated using a FEI Tecnai Supertwin F20 G2 operated at 200 kV. Site specific lift-outs for APT investigations were prepared from regions of the crept specimens (1% and 5% crept) containing a high dislocation density by using a dual-beam SEM/focused-ion-beam (FIB) instrument (FEI Helios Nanolab 600i). The adopted detailed procedure was described in ref.[46]. APT measurements were conducted on a Cameca LEAP 5000 XR operated in laser pulsing mode at a pulse repetition rate of 125 kHz and a pulse energy of 45 pJ. The specimen base temperature was kept at 60 K and the detection rate maintained at 1 ion detected per 100 pulses on average.

Figure 1 shows representative overviews of the microstructure, viewed along the [111] direction, for the samples crept up to 1% (1% sample) and 5% strain (5% sample), respectively. Figure 1(a) shows the STEM bright-field (BF) image for the 1% sample, where we observe dislocations mostly confined to the γ channels. For the 5% sample, Figure 1(b), we observe a significant increase in dislocation density in the microstructure, in both, γ-channels and γ' precipitates, consistent with what was reported in ref. [43]. Due to the higher dislocation density in the 5% sample, a STEM DF image is shown for better dislocation contrast in Figure 1b.

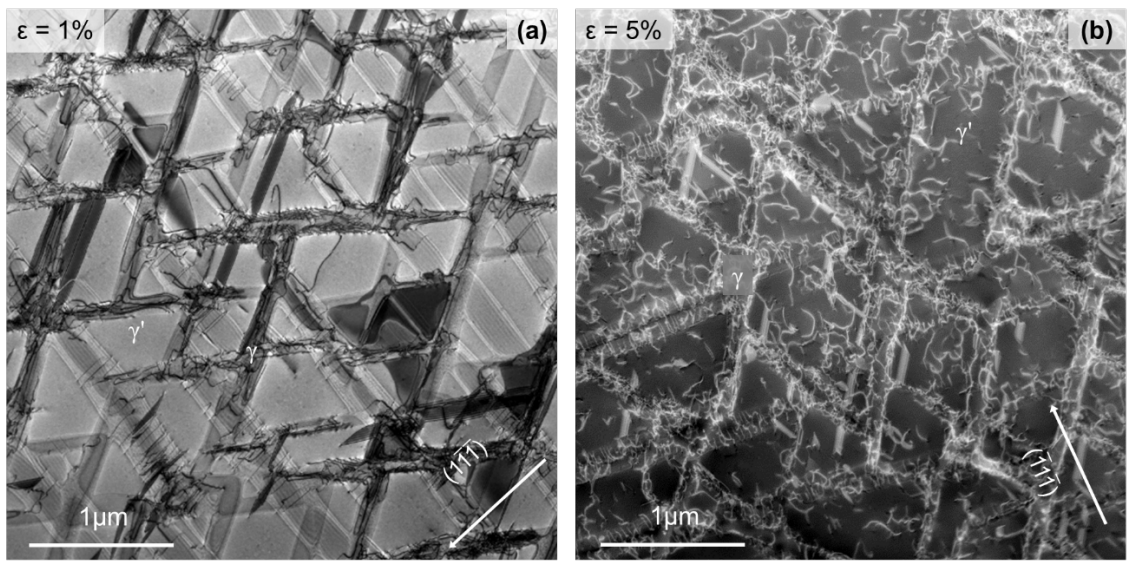

*Figure 1: STEM microstructure overview (a) Bright field (BF), 1% crept sample. (b) Dark field (DF), 5% crept sample.*

Figure 2(a) shows a 3D APT reconstruction from the 1% sample, showing the distribution of Co atoms colored in dark yellow. The γ/γ' interfaces are evidenced by the iso-composition surfaces delimiting regions containing more than 14 at.% Co. The reconstruction contains a γ channel with high Co-composition in between two γ' precipitates. Many tertiary γ' particles which formed during cooling can be seen in the γ channel [47]. The Co solute partitions to the γ phase. The compositional partitioning of different solutes across the γ/γ' interface is shown in supplementary Figure S1. The base component Ni and the solutes Al, Ti, Ta, and W show partitioning to the γ' precipitate while the other solutes (Co, Cr, Mo and Re) partition to γ matrix phase. This chemical partitioning of solutes is consistent with the literature [6,23,48].

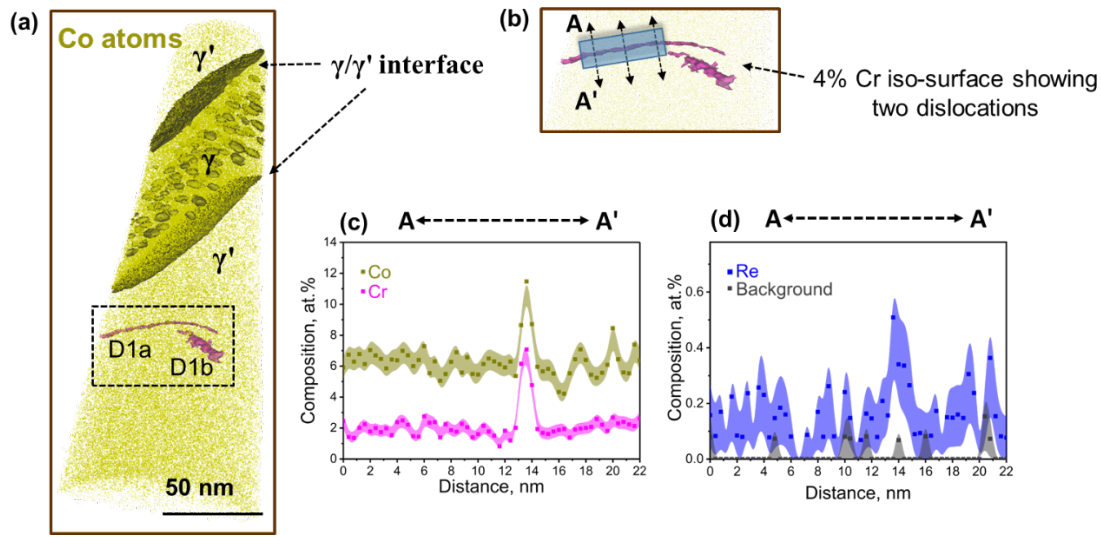

Figure 2: (a) APT reconstruction of a tip prepared from the 1% sample showing a γ-channel between two γ/γ' interfaces. Two dislocations D1a and D1b are highlighted in the dashed black rectangle region. (b) Magnified view for the dashed black rectangle, highlighting dislocations D1a and D1b. The composition profiles were recorded in the marked area. (c) Composition profile across dislocation D1a. Clear enrichments in Co and Cr are also observed. (d) Re composition profile in the same area. Re segregation across the dislocation is significant compared to the background level of Re.

Inside the γ' precipitate, the dashed black rectangle delineates a region containing two linear features decorated by Cr which correspond to two dislocations [49,50]. Figure 2(b) shows a magnified view of this region, with the two dislocations marked as D1a and D1b (D1 refers to dislocations observed in the 1% specimen). The composition profiles of Co, Cr and Re across D1a (i.e. along AA' from the blue cuboidal box in Figure 2(b)) are shown in Figures 2(c) and 2(d). We observe a significant enrichment of Co (by ~ +4 at.% ) and Cr (~ + 3.5 at.%) at D1a with respect to the surrounding γ' phase. Additionally, we find a relative increase in Re composition by ~ +0.4 at.%. The composition profiles of other solutes are plotted in supplementary Figure S2 which shows that Ni and Al are both depleted at dislocation D1a, alongside with a slight reduction of Ta and W. Ti and Hf show no significant change. A similar trend in the segregation of solutes is observed for dislocation D1b, as shown by the composition profiles provided in the supplementary Figure S3.

Figure 3 shows the APT analysis from the 5% sample. A similar reconstruction is shown in Figure 3(a), with the distribution of Co atoms and iso-compositional surfaces indicating a Co-composition of 14 at.% Co at the γ/γ' interfaces. An iso-surface surrounding regions containing more than 4 at.% Cr reveals three dislocations (D5a, D5b, D5c) inside the two γ' precipitates (similarly, D5 refers to dislocations in the 5% specimen). The region containing dislocation D5a is marked by a dashed black rectangle in Figure 3(a). Figure 3(b) is a close-up on dislocation D5a. A 2D compositional map of Re was projected onto the surfaces of a cuboid by keeping the region-of-interest at the center as shown in Figure 3(c), with the color-coded composition scale for solute Re. We observe a confined Re segregation along the dislocation line. Figure 3(d) shows the distribution of Re atoms (blue color) projected on the XZ and YZ planes (see the directions in Figure 3(c)). An iso-surface surrounding the region containing more than 1.5 at.% Re clearly encloses dislocation D5a with a high composition of Re atoms relative to the surrounding γ' lattice. Figures 3(e) and 3(f) show the composition profiles of Cr, Co, Re and Mo. We observe an enrichment of Re up to ~ 2.6 at.% at D5a that is significantly higher than the segregation observed at dislocations in the 1% sample. Additionally, a Mo enrichment up to approx. 1.3 at.% appears at D5a.

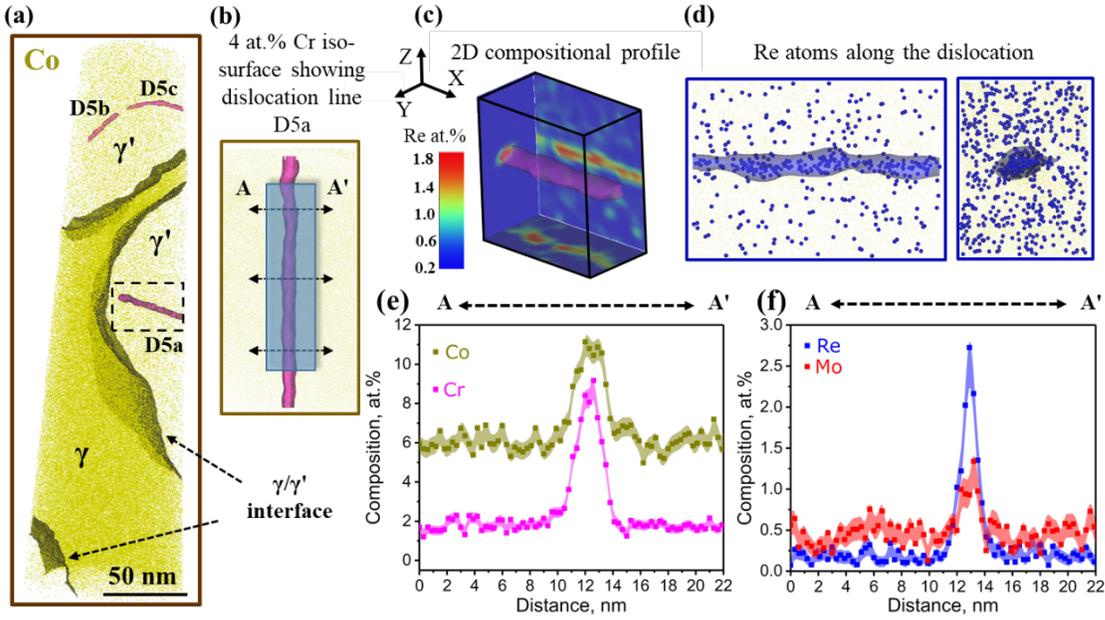

*Figure 3: (a) APT reconstruction of 5% deformation tip. Three dislocations are numbered as D5a, D5b and D5c. (b) Enlarged view of dislocation D5a. Composition profiles were measured in the marked region from A to A'. (c) 2D composition profile of dislocation 5a, projected onto the surfaces of a cuboid, showing the Re composition profile. (d) Re atom distribution on planes XZ and YZ, showing clear segregation of Re to dislocation D5a. (e) Composition profile of dislocation D5a, clear Co and Cr enrichment is observed at the dislocation. (f) Composition profile of Re and Mo in the same area as in (e), stronger segregation of Re can be seen in this condition compared to the 1% sample. Also, Mo is clearly enriched.*

The composition profile of Ni and other solutes are plotted in Figure supplementary S4. Composition profiles across dislocation D5c can be found in the supplementary Figure S5, those for dislocation D5b are not shown due to the limited number of atoms acquired. We observe similar solute segregation behavior as for dislocation D5a, however, the enrichment levels were different. We deduce that the segregation levels, particularly for solutes Re and Mo, increase as creep strain increases from 1 to 5%.

Referring to the creep curves for the two stages under investigation reported in ref. [43,44], it can be seen that the material exhibits significant primary creep. Previous studies on this material and similar CMSX-4 alloys indicate that dislocations form ribbons with an overall <112> fault vector inside of the γ' phase [45,51–53]. Solute decoration of dislocations, known as Cottrell atmospheres, have been reported in several alloy systems. The proposed mechanism is related to the solute interaction with the dislocation's elastic strain field which influences its mobility. In addition, several experimental studies have shown that solute atoms diffuse much faster along dislocation lines than in the bulk material. This effect is attributed to the reduced activation energies for diffusion in the dislocation core, an effect known as pipe diffusion [54–57]. The solute mass flux density depends on the "pipe cross-section" which is assumed as a disk with a radius of the size of the Burgers vector [54].

In γ/γ' based superalloys (both Ni- and Co-based), solute segregation was reported at dislocations present in the γ' precipitates [34,35]. All of them showed Co and Cr enrichment with respect to the surrounding γ' lattice which is also consistent with the results found for the present crept ERBO1 alloy. However, the mechanisms and driving forces for solute enrichments at dislocations are unclear. Earlier reports indicate that enrichment of Cr and Co is related to their smaller atomic size and higher mobility compared to other solutes such as W, Ti, and Mo [58]. This is in contrast to the present observation of additional Re and Mo (solutes partitioned in the γ phase) segregation confined to the γ' dislocations, since Re and Mo have larger atomic sizes and low diffusivities. To the best of our knowledge, Re or Mo segregation to dislocations inside γ' precipitates were so far not yet reported. Previous studies have shown that Re segregates to some dislocations which are at the γ/γ' interface[21,22], while Mo is observed to enrich along complex intrinsic stacking faults or superlattice intrinsic stacking faults in a single-crystal prototype nickel-based superalloy [38].

We now consider the mechanisms for the solute enrichment to the dislocations inside γ'. Figure 4(a) shows a schematic illustration of a γ' cuboidal precipitate embedded into a γ matrix with a dislocation moving on a {111} plane. Figure 4(b) shows the normalized composition profiles across the γ/γ' interface for the solutes partitioning to γ (γ stabilizers: Co, Cr, Re, and Mo) and γ' (γ' stabilizers: Ni, Al, W, Ta, and Ti). There are three possibilities of how solute enrichment at the dislocations can occur. (1) Local static rearrangement and long-range diffusion of solutes towards the dislocation core from the surrounding γ' lattice. (2) Solute collection in the γ' phase as the dislocation moves through the precipitate. (3) Diffusion of solutes from the γ matrix into the γ' phase along the dislocation line as depicted in the Figure 4(a). The compositional profile of the γ stabilizers in Figure 3(e-f) does not show any evidence for the presence of a compositional denuded zone surrounding the region enriched in solutes at the dislocation. In addition, the amount of γ stabilizing solute elements in γ' is very low, in particular for e.g. Re and Mo. Hence, possibilities (1) and (2) are unlikely to explain the high enrichment observed in the present work.

In turn, mechanism (3) can explain the solute segregation at the dislocations after creep in the γ' phase. For local thermodynamic equilibrium at the γ/γ' interface, the solutes should have equal chemical potentials in both γ and γ' phases, i.e. $\mu_i^\gamma = \mu_i^{\gamma'}$, where $i$ indicates the solute species. In the presence of a dislocation cutting into the γ'-phase and connecting the γ and the γ' phase, Figure 4(a), the solute exchange can be expected to take place along the dislocation line (mechanism (3)) having a well-defined pipe cross-section. The system tends to equilibrate the chemical potential of the solutes along the dislocation line. Hence, there is a driving force for a change of composition along the dislocation line. The composition of γ' stabilizers should decrease because of the less ordered structure of the dislocation core, bringing the composition locally closer to γ. In contrast, the composition of γ stabilizers increases along the dislocation core in γ'.

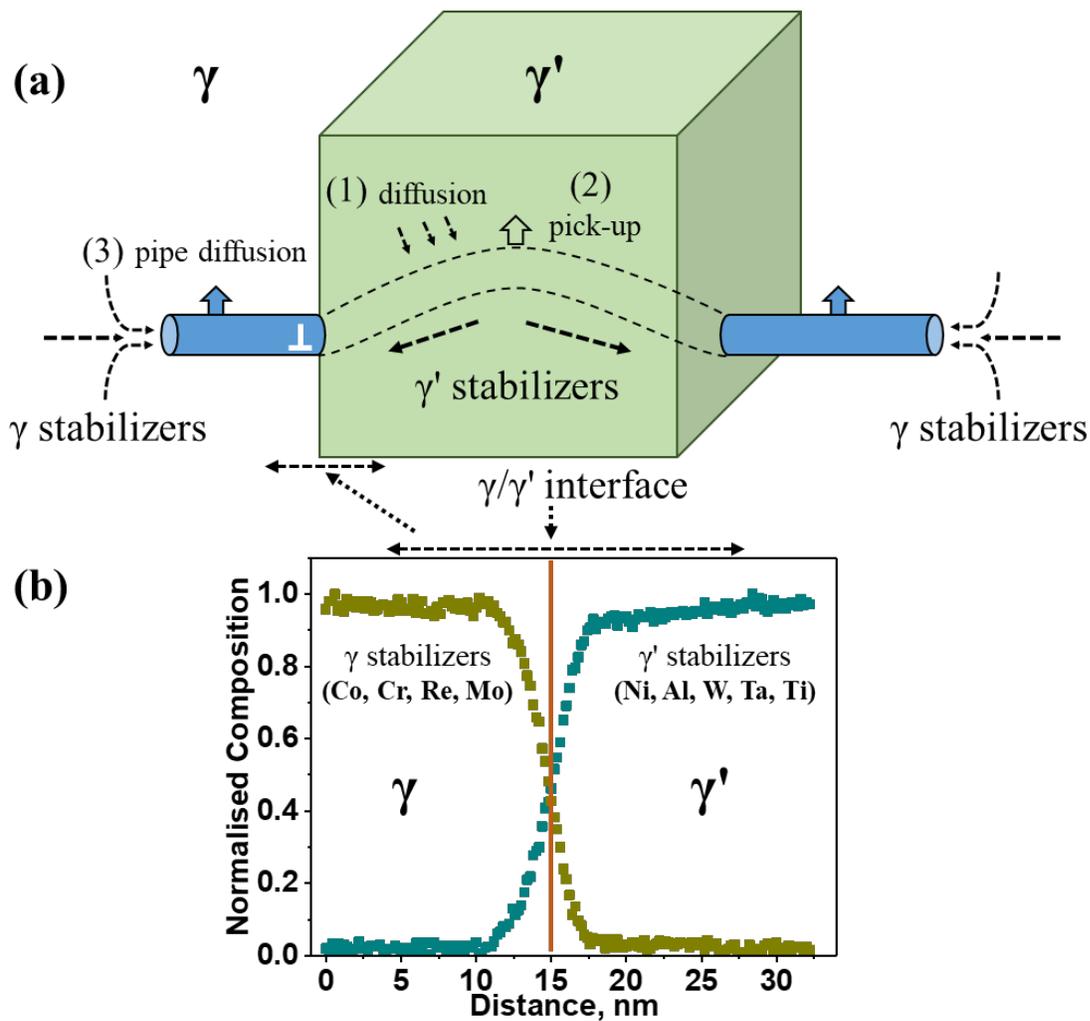

*Figure 4: (a) Schematic showing the three mechanisms of solute enrichment at the dislocation: (1) Long range diffusion of solutes from the surrounding γ' lattice, (2) Solute pick-up from the γ' lattice during the movement of the dislocation, and (3) Solute exchange through the dislocation line core between the γ and γ' precipitate. (b) Normalized composition profiles for γ and γ' stabilizers across the γ/γ' interface.*

The composition profiles in Figures 2(c) and 3(e) show a significant enrichment of Co and Cr at the dislocation inside the γ' precipitate for both, 1% and 5% samples. This result is consistent with the suggestion introduced above. We also observed a confined depletion of Ni and Al at the dislocation line within the γ' precipitate. Additionally, in the 5% samples, the pronounced segregation of Re and Mo, in addition to the enrichment of Co and Cr, to the dislocation also supports the proposed mechanism. Correspondingly, there is always a reduction of Ta and in some cases also of W and Ti (supplementary Figures S3, 4 and 5) at the dislocations. Transport of these solutes along the dislocation lines occurs at high temperatures (here: 750°C) also in the presence of an external load. Hence, the composition of the dislocations in the γ' precipitates is expected to also depend on the creep strain and may vary along the line of a dislocation. In the present study, we generally observe that for the 5% sample (172h creep exposure), the segregation of Co, Cr, Re and Mo is significantly more pronounced as compared to the 1% specimen (3h creep exposure). This shows that with increasing creep exposure time at 750°C, more diffusion takes place and hence a higher solute content can accumulate at the dislocations.

In summary, the current APT results provide experimental evidence for Re and Mo segregation to the dislocations present inside of γ' precipitates. The results shown here can be explained in terms of a pipe diffusion mechanism, where solute mass transport occurs along the dislocation line out of the γ matrix into the γ' phase, thereby crossing the γ/γ' interface. This links the field of γ/γ' dislocation plasticity to partitioning and segregation processes in γ/γ' microstructures. The suggested process can influence the creep behavior as solute decoration of dislocations creates a solute drag force which slows down dislocation movement. Moreover, the selective solute transport along the dislocations in the γ' phase is in accordance with mechanism (3) which depends on the solute partitioning behavior across interfaces. Hence, the change in solute partitioning behavior will also affect the type and direction of the movement of solutes along the shearing dislocations across the γ/γ' interfaces. Control over the solute partitioning is possible by tuning alloy chemistry. For example, addition of Ta to a Ni-based superalloys has been shown to reverse the partitioning of W from the γ' phase into the γ precipitate [59]. Similarly, in Co-based superalloys, Cr addition leads to a similar effect on the Mo partitioning behavior [60]. These observations can serve as guiding principles for devising improved alloy doping concepts for high temperature applications.

## ACKNOWLEDGMENTS

The authors are grateful to U. Tezins and A. Sturm for their technical support of the atom probe tomography and focused ion beam facilities at the Max-Planck-Institut für Eisenforschung. XW, SKM, BG, DR and GE are grateful for funding from the **Deutsche Forschungsgemeinschaft (DFG)** through **Project A2 and A4** of the collaborative research center SFB/Transregio 103 superalloy single crystals under grant number **INST 213/747-2**. SKM also acknowledges financial support from the **Alexander von Humboldt Foundation**. BG was an editor of *Materialia* at the time this article was submitted and accepted but was not involved in the peer-review process or in making the decision on this manuscript. BG acknowledges financial support from the **ERC-CoG-SHINE-771602**.

**Declaration of interest**

The authors have no potential conflict of interest to disclose.

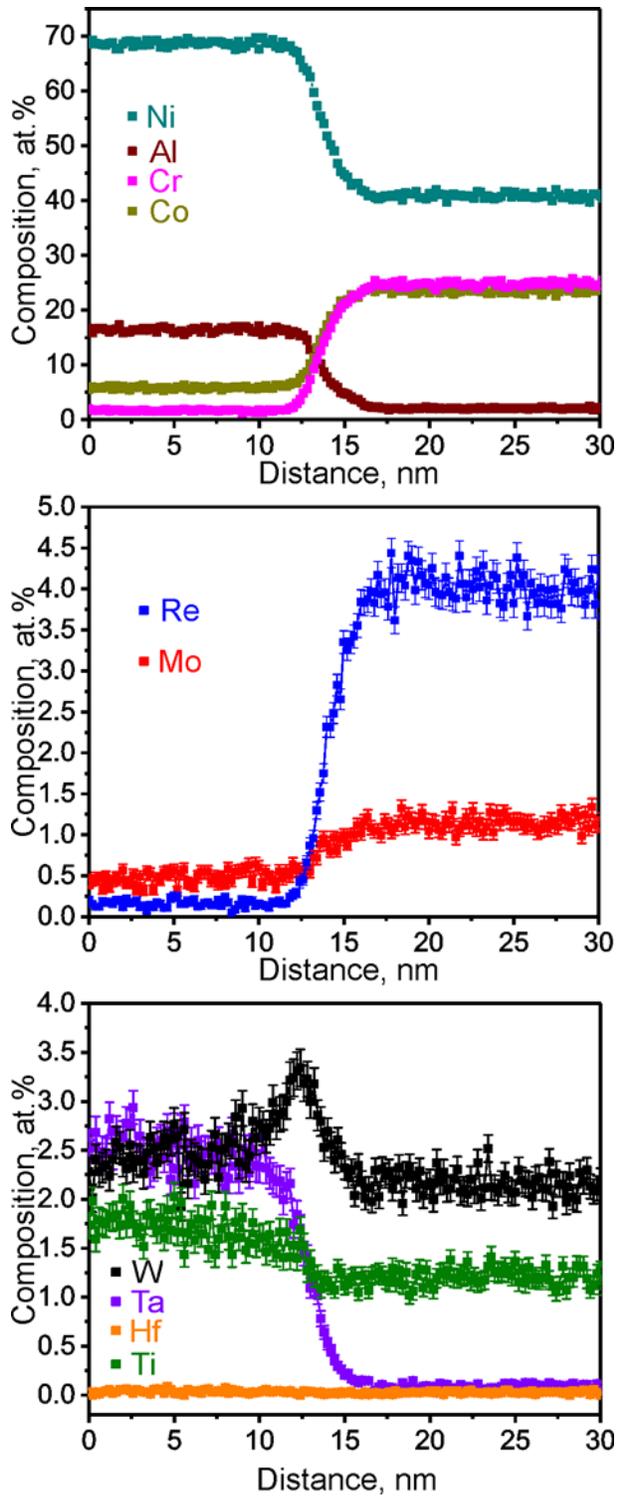

*Figure S1: Composition profiles across γ/γ' interface at 1% creep. (a)Distribution of Ni, Al, Co and Cr across γ/γ' interface. Ni and Al are enriched in the γ' phase, and Co and Cr are enriched in the γ phase. (b) Distribution of Re and Mo across γ/γ' interface. Both elements are enriched in the γ phase. (c) Distribution of W, Ta, Hf and Ti across*

*γ/γ' interface. W and Ta are enriched in the γ' phase. Ti is slightly enriched in the γ' phase. Hf shows no obvious enrichment behavior.*

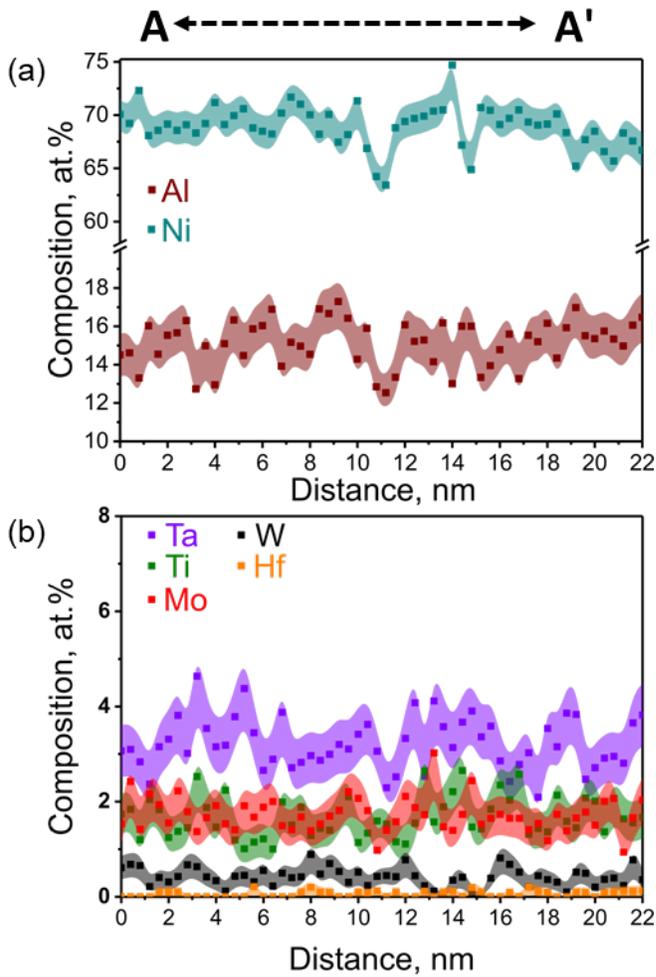

*Figure S2: Compositional profile of the same region for dislocation D1a in 1% creep strain tip. (a) Distribution of Ni and Al across dislocation D1a. Clear depletion of Ni and Al can be seen. (b) Distribution of Ta, Ti, W, Hf and Mo across dislocation D1a. No profound change of composition of these elements are observed.*

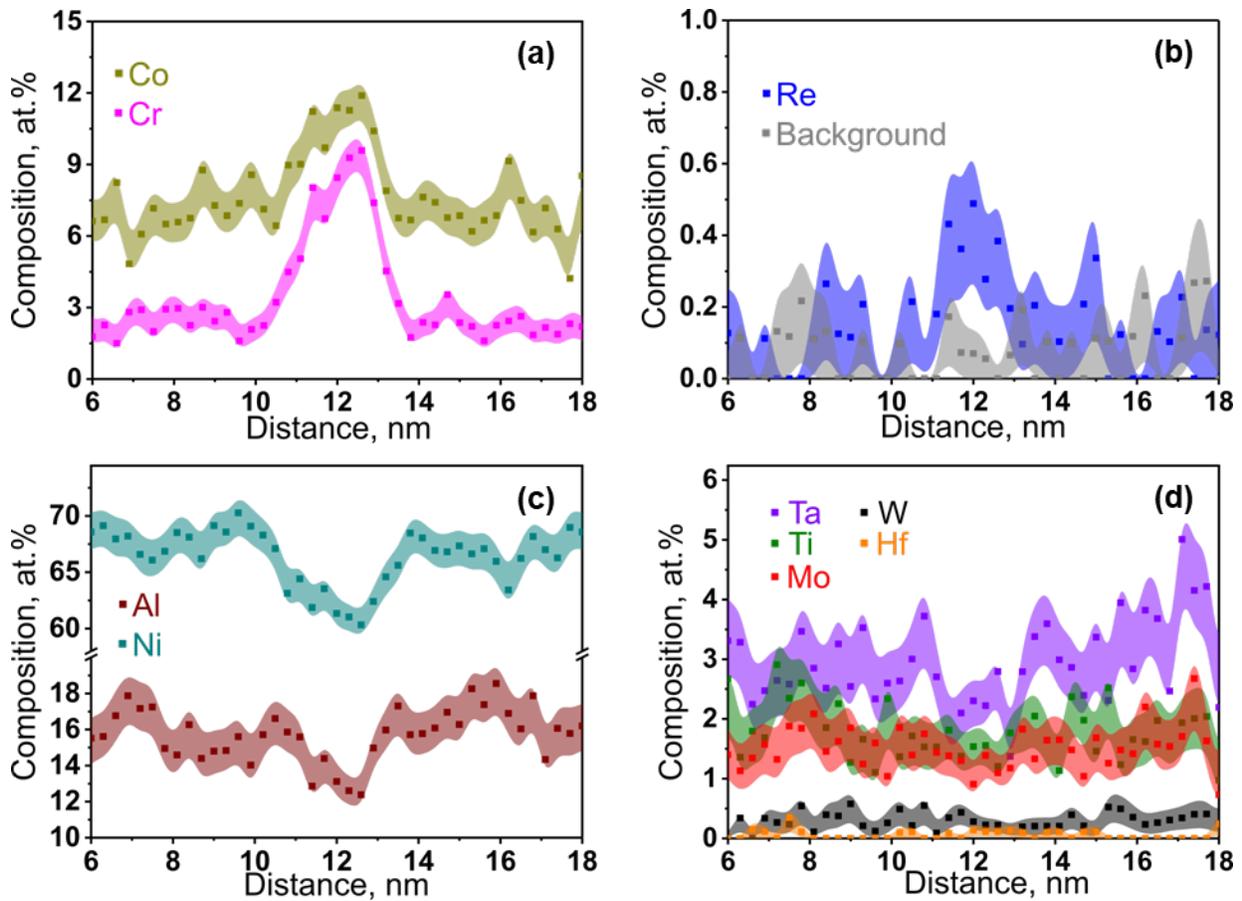

*Figure S3: Compositional profile of cube investigation area for dislocation D1b in 1% creep strain tip. (a) distribution of Cr and Co across dislocation D1b. Clear enrichment of Co and Cr can be seen. (b) Distribution of Re across dislocation D1b. There is nearly 0.6 at. % of Re at the dislocation, compared to an average of 0.2 at.% for the surrounding γ' phase. (c) Distribution of Ni and Al across the dislocation, clear depletion of Ni and Al is observed for the dislocation region. (d) Distribution of Ta, Ti, Mo, W and Hf across dislocation. Clear depletion of Ta is observed for the dislocation, while for other elements there are no profound changes.*

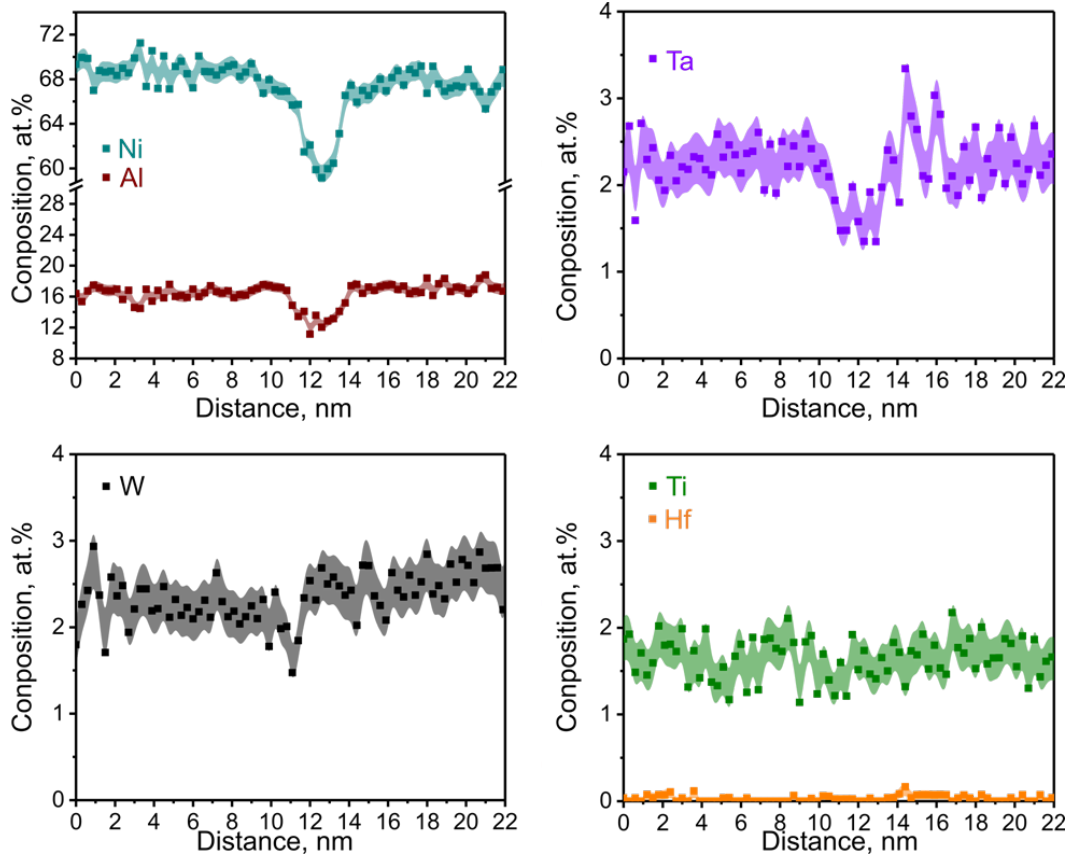

*Figure S4: Compositional profile of dislocation D5a in 5% creep strain tip, same investigated area as in Figure 3. (a) Distribution of Ni and Al across the dislocation, clear depletion of Ni and Al is observed for the dislocation region. (b) Distribution of W, Ti, Ta and Hf across dislocation. Clear depletion of Ta is observe for the dislocation, as well as W. For Ti and Hf there are no profound changes.*

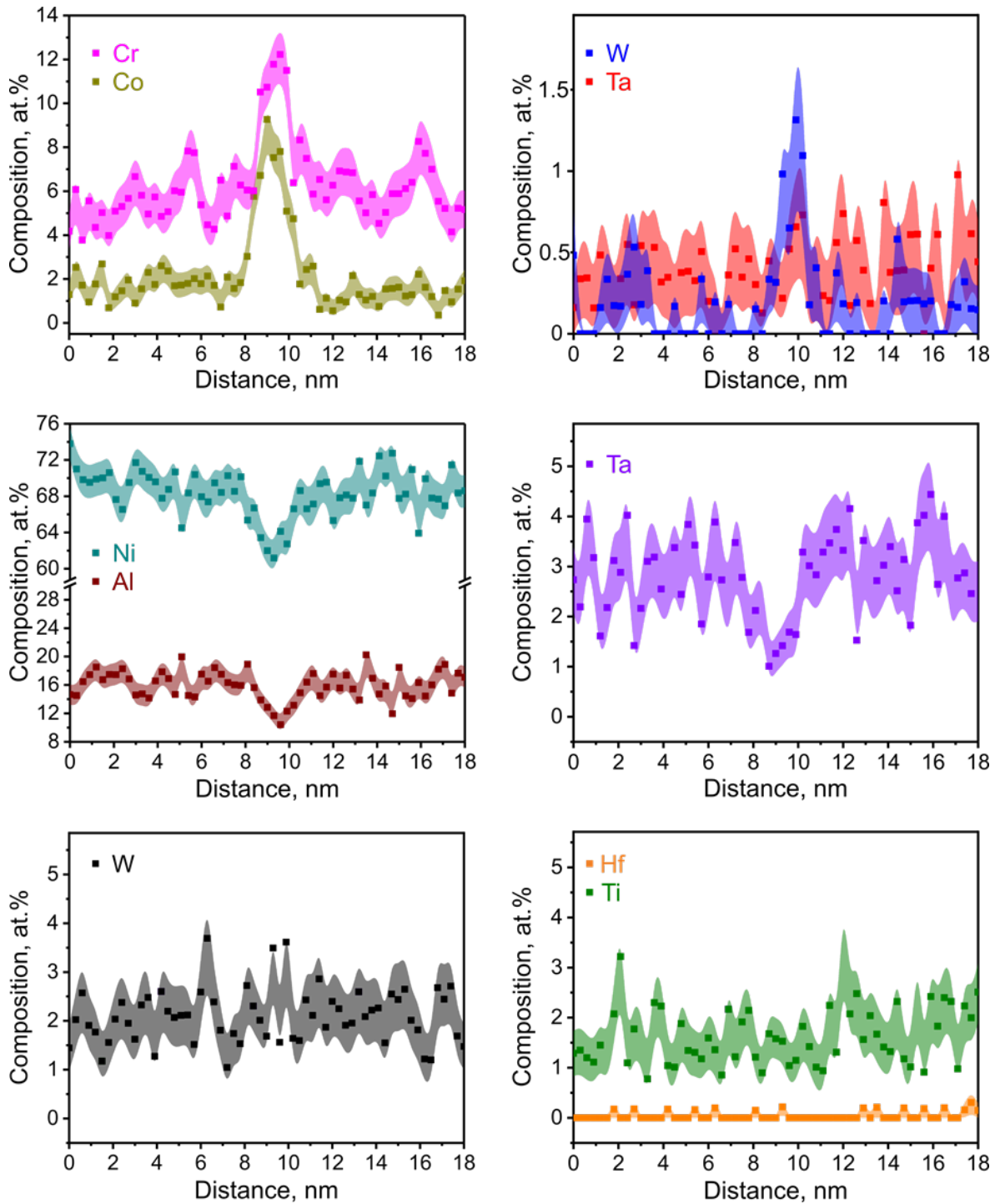

*Figure S5: Compositional profile for dislocation D5c in 5% creep strain tip. (a) distribution of Cr and Co across dislocation D5c. Clear enrichment of Co and Cr can be seen. (b) distribution of Re and Mo across dislocation D5c. There is nearly 1.5 at. % of Re at the dislocation, compared to an average of 0.4 at.% for the surrounding γ' phase. (c) distribution of Ni and Al across the dislocation, clear depletion of Ni and Al is observed for the dislocation region.*

*(d) distribution of Ta, Ti, Mo, W and Hf across dislocation. Clear depletion of Ta is observed for the dislocation, while for other elements there are no profound changes.*